\documentclass[prl,reprint, notitlepage, superscriptaddress,longbibliography]{revtex4-2}
\usepackage{braket}
\usepackage{graphicx}
\usepackage{amsmath}
\usepackage{amssymb}
\usepackage{hyperref}
\usepackage[all]{hypcap}
\hypersetup{colorlinks=True,allcolors=blue}
\usepackage{xcolor}
\usepackage{subfigure}
\usepackage{placeins}
\usepackage{times}
\usepackage{bbold}
\DeclareMathOperator{\Tr}{Tr}

\begin{document}
\author{Xin H. H. Zhang}
\email[Corresponding author: ]{physicsxinzhang@gmail.com}
\affiliation{Technical University of Munich, TUM School of Natural Sciences, Physics Department, 85748 Garching, Germany}
\affiliation{Walther-Mei{\ss}ner-Institut, Bayerische Akademie der Wissenschaften, 85748 Garching, Germany}
\affiliation{Munich Center for Quantum Science and Technology (MCQST), 80799 Munich, Germany}

\author{Daniel Malz}
\affiliation{Department of Mathematical Sciences, University of Copenhagen, Universitetsparken 5, 2100 Copenhagen, Denmark}

\author{Peter Rabl}
\affiliation{Technical University of Munich, TUM School of Natural Sciences, Physics Department, 85748 Garching, Germany}
\affiliation{Walther-Mei{\ss}ner-Institut, Bayerische Akademie der Wissenschaften, 85748 Garching, Germany}
\affiliation{Munich Center for Quantum Science and Technology (MCQST), 80799 Munich, Germany}

\date{May 19, 2025}

\title{Unraveling superradiance: Entanglement and mutual information in collective decay}

\begin{abstract}
We study the collective decay of an initially inverted ensemble of two-level emitters in two distinct scenarios: when coupled to a squeezed photonic reservoir and when interacting with a one-dimensional waveguide. Using a quantum-state diffusion approach to unravel the emission process, we investigate entanglement and classical correlations along individual quantum trajectories over time. This numerical analysis shows that despite an initial build-up of entanglement and a significant amount of entanglement due to either spin squeezing or dark states at late times, the essential features of the superradiant burst are well described by averages over fully factorizable states. We explain this observation in terms of an almost complete factorization of all 2-local observables, which we identify as a generic property of superradiant decay. Based on this insight, we provide a purely classical theory for the burst in squeezed superradiance, which is both intuitive and exact for a large number of emitters. Moreover, we find that our numerical approach also performs well in the presence of subradiant states, which dominate the slow residual decay of non-uniform ensembles at late times. 
\end{abstract}

\maketitle

Superradiance is a prototypical example that shows how the build-up of environment-induced correlations can drastically accelerate the decay of a large ensemble of two-level emitters~\cite{DickePhysRev1954,GrossPhysRep1982}. Interestingly, this phenomenon can be explained either in terms of transitions between highly entangled Dicke states~\cite{DickePhysRev1954}, or as a dissipative evolution of a mixture of spin coherent states~\cite{NarducciPRA1974a}. While the former interpretation suggests that quantum entanglement plays an important role in superradiance, in the latter description, the emission burst can be described classically, but seeded by initial quantum fluctuations. In conventional Dicke superradiance, where the underlying permutation symmetry permits precise analytic predictions~\cite{RehlerPRA1971,BonifacioPRA1971,BonifacioPRA1971b,DegiorgioOC1971,DegiorgioPRA1971,HaakePRA1972,NarducciPRA1974,NarducciPRA1974a,Agarwal1974,GrossPhysRep1982,MalzPRA2022,holzinger2025},
both descriptions lead to the same result~\cite{NarducciPRA1974a}.
However, in more general situations without permutation symmetry, the situation is much less clear, and the question whether entanglement is necessary to describe the physical process gains immediate relevance, because numerical simulations are much more efficient in the absence of entanglement. This becomes particularly important for the study of subradiant states~\cite{PrasadPRA2000,GuerinPRL2016,AsenjoGarciaPRA2017,AsenjoGarciaPRX2017,AlbrechtNJP2019,HolzingerPRL2022,FasserOQ24,TiranovScience2023,SheremetRMP2023,HenrietPRA2019,KornovanPRA2019,WangPRL2020,ZhangPRL2020,PoshakinskiyPRL2021,GlicensteinOL2022,RubiesBigordaPRA2022,PineiroPRX2022,SantosPRA2022,RubiesBigordaPRR2023,FasserOQ24,ShiPRA2024,FerioliPRL2025,LeeArxiv2025},
which determine the long-time residual decay of non-uniform ensembles, and are expected to be highly entangled.

In this Letter, we study this question in two scenarios beyond Dicke superradiance. First, we address the decay of a permutation-invariant ensemble of two-level systems (TLSs) into a squeezed reservoir, and second, we consider an array of TLSs with spatially varying couplings to a one-dimensional (1D) reservoir. In the former scenario, it is well-known that the emitters relax into a steady state with a large degree of spin squeezing~\cite{AgarwalOC1989,AgarwalPRA1990,DallaTorrePRL2013,MunozArxiv2019,GroszkowskiPRX2022,KoppenhoferPRL2023,BarberenaPRA2024},
which provides a direct witness for entanglement~\cite{SorensenNature2001}.
Thus the natural question arises of how these quantum correlations emerge during the decay process and how they affect the superradiant burst. To address this question, here we use a quantum state diffusion (QSD) approach~\cite{GisinJPA1992,Carmichael1993} to unravel the evolution of the mixed state of the system in terms of individual quantum trajectories~\cite{Dalibard1992,Gardiner1992,GisinJPA1992,TeichPRA1992,Carmichael1993}. Along each stochastic trajectory, the system is in a pure state and can be represented as matrix product state (MPS)~\cite{SchollwockRMP2005,CiracRMP2021}, with an adjustable degree of entanglement. While from these simulations we observe a superradiant growth of entanglement already in the initial phases, these quantum correlations do not affect the central features of superradiance, which are accurately reproduced by averages over fully factorized states. As an immediate consequence, this finding allows us to derive an intuitive analytic model, which accurately predicts all the features of the emission burst for squeezed superradiance.     

\begin{figure}\label{Fig_PopRate}
	\centering	\includegraphics[width=\columnwidth]{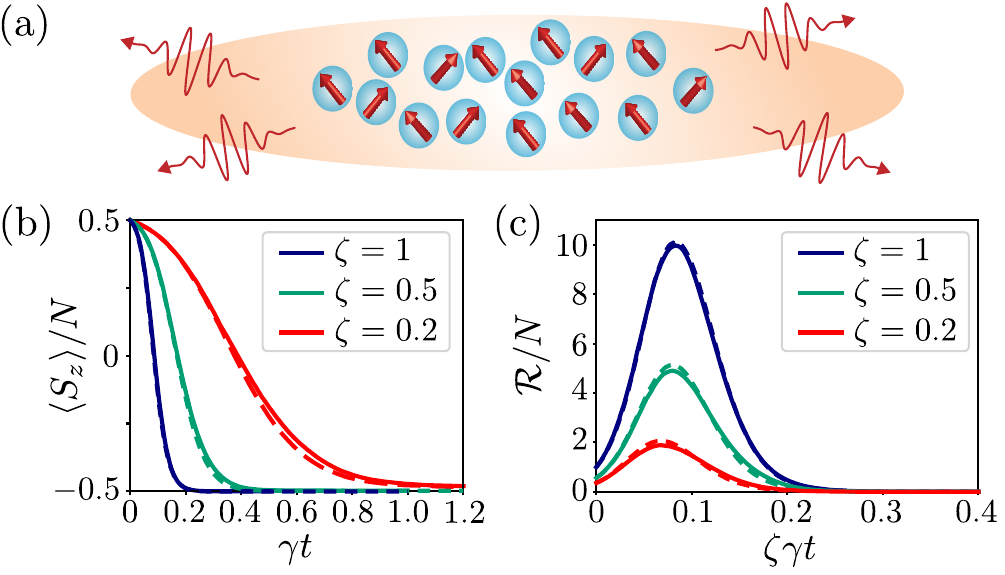}
	\caption{ (a) Schematic of an ensemble of two-level dipoles emitting collectively into a squeezed reservoir. Time evolution of (b) the magnetization $\braket{S_{z}}/N$, and (c) the average decay rate $\mathcal{R}/N$, as obtained from QSD simulations with $N=50$ TLSs and a bond dimension of $d=10$. For comparison, the results with $d=1$ are shown by the dashed lines.  
    For these plots, we used a step size of $\delta t = 2\times 10^{-4}(\gamma\zeta)^{-1}$ and averaged over $N_{\text{tr}} =10^{3}$ trajectories.}
\end{figure}

In the second scenario, of a 1D array of TLSs with spatially varying couplings, it is known that the system occupies entangled dark states at late times after the superradiant burst.
However, also in this case we find that the factorized state accurately reproduces the exact decay, including the late-time dynamics dominated by subradiant states. We explain this agreement by the suppression the total mutual information between any two TLSs along each trajectory. Therefore, while significant entanglement can be generated during both superradiant and subradiant decay, our results show that many relevant observables can be computed without taking such quantum correlations into account. This observation suggests that QSD simulations with factorizable states are already sufficient to model collective radiation processes in large ensembles with spatial varying couplings and in the presence of squeezing---scenarios that have been out of reach of efficient numerical simulations so far.

\emph{Squeezed Superradiance.}---The collective radiative decay of $N$ TLSs into a homogeneous, squeezed reservoir can be modeled by a master equation of the form~\cite{AgarwalOC1989,AgarwalPRA1990}
\begin{equation}\label{SSR_ME}
\frac{d}{dt} \rho  = \gamma \mathcal{D}[S_x - i \zeta S_y] \rho,     
\end{equation}
where $\rho$ is the system density operator and $\gamma$ the decay rate of a single TLS. In Eq.~\eqref{SSR_ME} we have defined $\mathcal{D}[J] \rho = J \rho J^{\dagger} - \frac{1}{2} \{ \rho, J^{\dagger}J \}$ and introduced the collective spin operators $S_{\alpha} = \sum_{j=1}^{N} \sigma_{j}^{\alpha}/2$ with $\sigma_{j}^{\alpha}$ denoting the usual Pauli operators. The parameter $\zeta\in [0, 1]$ quantifies the degree of squeezing, where for the chosen parametrization~\cite{BarberenaPRA2024,SupMat} we recover the case of conventional Dicke superradiance when $\zeta=1$, while $\zeta=0$ corresponds to the extreme limit of an infinitely squeezed reservoir. For intermediate $0<\zeta<1$ and $\zeta N > 1$, the steady state of Eq.~\eqref{SSR_ME} is a spin-squeezed state~\cite{AgarwalOC1989,AgarwalPRA1990,DallaTorrePRL2013,MunozArxiv2019,GroszkowskiPRX2022,KoppenhoferPRL2023,BarberenaPRA2024} with spin-squeezing parameter $\xi_{\text{R}}^{2} \approx \zeta$. Note that a value of $\xi_{\text{R}}^{2}<1$ certifies that the state is entangled~\cite{SorensenNature2001}. In the following, we are primarily interested in the decay dynamics into this state, starting from all TLS initially excited, $|\psi_0\rangle=|1\rangle^{\otimes N}$.

\emph{Quantum trajectories.}---While Eq.~\eqref{SSR_ME} is permutation invariant and can be solved numerically for rather large $N$, here we proceed with a QSD unraveling~\cite{GisinJPA1992,Carmichael1993,VerstraelenPRXQ2023,LiArxiv2025,SantiniPRB2025} of this master equation, which describes the system's dynamics under homodyne measurement~\cite{WisemanMilburn2009}. In this approach, the evolution of the density operator $\rho$ is calculated from an average over pure states $|\psi\rangle$, each of which evolves according to a nonlinear stochastic Schr\"odinger equation of the form 
\begin{align}\label{QSDgeneral}
    d \ket{\psi} = & - \frac{1}{2}  \left( J^{\dagger} J + \braket{J}^{*} \braket{J} - 2 \braket{J}^{*} J \right) \ket{\psi} dt \nonumber \\ &+ (J - \braket{J}) \ket{\psi} dW -i H \ket{\psi} dt.
\end{align}
Here $dW$ is a complex Wiener increment and in our current example $J=\sqrt{\gamma}(S_x-i\zeta S_y)$ and $H=0$. A generalization with multiple jump-operators and a non-vanishing Hamiltonian is discussed below. Along each trajectory, we represent the state $|\psi\rangle$ as a MPS~\cite{SchollwockRMP2005,CiracRMP2021} with bond dimension $d$. 
This representation becomes particularly efficient when entanglement is low and it allows us to extend our simulations to rather large numbers of TLSs.  From a sufficiently large set of quantum trajectories that are simulated in this way, $\{ |\psi_s\rangle|s=1,\dots,N_{\rm tr} \}$, expectation values of observables can be computed as $\braket{\hat{O}} \simeq \sum_{s} \braket{\psi_s | \hat{O} | \psi_{s} }/N_{\rm tr}$. Note that unlike uncontrolled approximation methods, here numerically exact results can be approached by systematically increasing the bond dimension $d$ and the number of trajectories $N_{\rm tr}$.

\begin{figure}[tp]\label{Fig_Entropy}
	\centering	\includegraphics[width=\columnwidth]{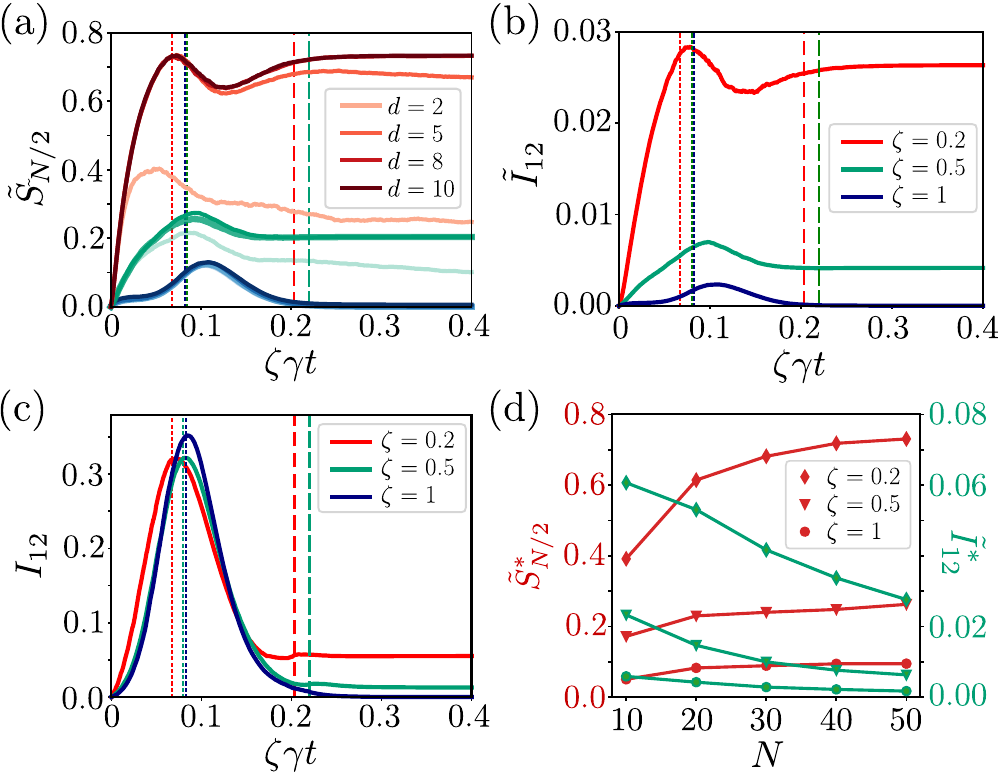}
	\caption{Summary of the results obtained from QSD simulations for $N=50$ and different squeezing parameters $\zeta$. (a) Plot of the average entanglement entropy $\tilde{S}_{N/2}$ for increasing bond dimension $d=2,5,8,10$. For all $\zeta>0.2$ the simulations converge for $d\geq10$. (b) Plot of the average mutual information between a pair of TLSs, $\tilde{I}_{12}$, for $d=10$. (c) Evolution of the  mutual information $I_{12}$ between a pair of TLSs without unraveling. (d) Dependence of the average entanglement $\tilde{S}_{N/2}$ and average mutual information $\tilde{I}_{12}$ at the peak time $t^{*}$ on the number of TLSs, $N$. In (a)-(c) the vertical dotted lines indicate $t^*$ and the vertical dashed lines the time beyond which $\xi_{\rm R}^2<1$. For all simulations we used $N_{\text{tr}} =10^{3}$ trajectories and a step size of $\delta t = 2\times 10^{-4} (\gamma\zeta)^{-1}$.}
\end{figure}

\emph{Entanglement and mutual information.}---In Fig.\,\ref{Fig_PopRate}, we use the QSD method to evaluate the evolution of the average magnetization, $\langle S_z\rangle/N$, and the instantaneous emission rate, $\mathcal{R}(t)=- \frac{d}{dt}\langle S_{z}\rangle(t)$, for different squeezing strengths.  We see that independent of the squeezing strength, the decay is collectively accelerated and a characteristic superradiant peak develops as correlations between the TLSs build up. More squeezing, however, leads to a smaller maximal rate $\mathcal{R}^{*}$, and the emission also peaks at a later time $t^{*}$. Surprisingly, even for $\zeta<1$, the evolution of $\langle S_z\rangle$ and $\mathcal{R}$ is already accurately reproduced for a bond dimension of $d=1$, which corresponds to a mean-field (MF) factorization of each trajectory. This seems contradictory to the fact that the spins eventually decay into an entangled, spin-squeezed state.

To obtain additional insights in the role of quantum and classical correlations during superradiant decay, we compute the average entanglement entropy $\tilde{S}_{\text{A}\bar{\text{A}}} = \sum_{s}  S_{\text{vN}}(\rho_{s,\text{A}})/N_{\rm tr}$ for a given bipartition $A\bar A$ of the ensemble. Here $S_{\text{vN}}(\rho_{s,\text{A}}) = -\Tr\{\rho_{s,\text{A}} \log_{2}(\rho_{s,\text{A}})\} $ is the von Neumann entropy and $\rho_{s,\text{A}} = \Tr_{\bar{\text{A}}} \{\ket{\psi_s} \bra{\psi_s}\} $ is the reduced density matrix of subsystem $\text{A}$ for a given quantum trajectory $\ket{\psi_s}$. Note that $\tilde{S}_{\text{A}\bar{\text{A}}}$ differs from the usual entropy evaluated from the full state of the system \emph{after} averaging over trajectories, $\rho= \sum_{s} \ket{\psi_s} \bra{\psi_s} /N_{\rm tr}$. It quantifies, however, more accurately the typical amount of entanglement that is present in each trajectory, as relevant for numerical simulations.  In Fig.~\ref{Fig_Entropy}(a) we plot $\tilde{S}_{N/2}$ for a balanced bi-partitioning of the system. We see that the average entanglement increases rapidly during the build-up of the superradiant burst, and reaches a maximum roughly at the time of the peak, $t^*\sim 1/(\zeta \gamma N)$. For $\zeta<1$, the average entanglement then remains approximately the same from this time on. For conventional superradiance with $\zeta=1$, a small amount of entanglement is present only in the vicinity of the peak. 

We emphasize that the average entanglement is only an upper bound for the entanglement of formation $S_{\text{F}}$ and that a nonvanishing value of $\tilde{S}_{N/2}$ does not necessarily imply that the state $\rho$ is entangled. However, we can also evaluate the spin squeezing parameter $\xi_{\text{R}}^{2}$ and use it as an independent entanglement witness. In Fig.\,\ref{Fig_Entropy}, the time after which the systems reaches a state with $\xi_{\text{R}}^{2} <1$ (see also End Matter) is marked by the dashed vertical lines. Beyond this line, we obtain nonvanishing values for both $\tilde{S}_{N/2}$ and $S_{\text{F}}$. Although there is no one-to-one correspondence, this observation suggests that $\tilde{S}$ can serve as qualitative estimate of the amount of entanglement in superradiance and that it is not just an artifact of our chosen unraveling.

Another insightful quantity is the average mutual information, which for two subsystems A and B is defined as $\tilde{I}_{\text{AB}}  = \sum_{s} [S_{\text{vN}} (\rho_{s,\text{A}}) + S_{\text{vN}} (\rho_{s,\text{B}}) - S_{\text{vN}} (\rho_{s,\text{AB}}) ]/N_{\rm tr}$. Here again, the states $\rho_{s,\text{X}}$ denote the respective reduced density operators derived from a single trajectory, $ \rho_s=|\psi_s\rangle\langle \psi_s|$. This quantity can be compared to the mutual information for the trajectory-averaged density operator $\rho$, $I_{\text{AB}}  = S_{\text{vN}} (\rho_{\text{A}}) + S_{\text{vN}} (\rho_{\text{B}}) - S_{\text{vN}} (\rho_{\text{AB}})$. In Fig.\,\ref{Fig_Entropy}(b) and (c) we plot the evolution of both quantities for the case where subsystems $A$ and $B$ represent a single TLS each. Interestingly, the mutual information of two spins in the trajectories $\tilde I_\text{AB}$ is more than an order of magnitude smaller than the mutual information in the overall mixed state $I_\text{AB}$. The difference is that $I_{12}$ additionally captures classical correlations within the full ensemble of trajectories, which follow the trend of the emission rate. We conclude that, while entanglement is clearly present in the trajectories for squeezed superradiance, it does not significantly influence the features of the superradiant burst, which instead arises primarily from emerging classical correlations between the radiating dipoles. 

In Fig.\,\ref{Fig_Entropy}(d), we repeat these simulations for different system sizes and evaluate the resulting average entanglement and average mutual information at the peak emission time $t^*$. From this plot, we find that the peak of the average entanglement $\tilde{S}_{N/2}$ follows an area law, and remains approximately constant despite the underlying all-to-all interactions. More relevant, the average mutual information between two TLSs decreases monotonically as $N$ increases. This means that in the relevant limit of large $N$, the state of any two TLSs factorizes and, after unraveling using QSD, we can write the 2-local reduced density matrices as
\begin{equation}\label{unravelrho12}
    \rho_{jj^{\prime}} = \frac{1}{N_{\text{tr}}} \sum_{s}  \rho_{s, jj^{\prime}} \approx \frac{1}{N_{\text{tr}}} \sum_{s}  \rho_{s, j} \otimes \rho_{s, j^{\prime}}.
\end{equation}
Note that this statement goes beyond the quantum de Finetti theorem \cite{Hudson1976,CavesJMP2002}, which states that for large systems, permutation-invariant states have separable local reduced density matrices. Here, after unraveling, the trajectories are under a stronger condition---their local reduced density matrices, such as $\rho_{s,12}$, are not just separable but product states and therefore also classically uncorrelated. This fact explains why 2-local operators, such as the decay rate with $\hat{\mathcal{R}} = \frac{\gamma}{4} \sum_{j}[ (\zeta^2+1) \sigma_{j}^{z} + 2 \zeta ] + \frac{\gamma}{4} \sum_{j}\sum_{l\neq j} \zeta(\sigma_{j}^{x} \sigma_{l}^{x} + \sigma_{j}^{y} \sigma_{l}^{y} )$, can be accurately reproduced by fully factorized quantum trajectories with $d=1$. This does not contradict the existence of spin-squeezing during and after decay, which is determined by non-extensive corrections of the spin variances caused by residual corrections to Eq.~\eqref{unravelrho12} (see End Matter).

\emph{An analytic theory for squeezed superradiance.}---Based on these numerical findings, we conclude that superradiance can be accurately modeled in terms of ensembles of fully factorized states. Thus, taking also permutation symmetry into account, we can make the ansatz $\ket{\psi(t)}_{s}= \prod_j |\theta(t),\phi(t)\rangle_j$ for each trajectory, where $|\theta,\phi\rangle= \cos(\theta/2)|1\rangle+ e^{i\phi}\sin(\theta/2)|0\rangle$. By plugging this ansatz into Eq.~\eqref{QSDgeneral} we can derive equivalent stochastic equations for the polar and azimuthal angles $\theta$ and $\phi$. These equations are summarized in End Matter and, after addition approximations valid in the limit $\zeta N\gg1$, we obtain the stochastic differential equation
\begin{equation}\label{ThetaEOMmain}
    d\theta \approx  - \partial_\theta V(\theta) dt + \sqrt{\frac{\gamma}{2}} dU(t),
\end{equation}
which in this limit is also decoupled from the evolution of $\phi$. In Eq.~\eqref{ThetaEOMmain}
we have introduced the effective potential 
\begin{equation}\label{Potential}
    V(\theta) \approx \frac{\gamma}{2} \zeta N  \cos \theta - \frac{\gamma}{4} (1+\zeta)^2 \ln\theta - \frac{\gamma}{4} (1-\zeta)^2 \ln(\pi-\theta),
\end{equation}
and $dU(t) \approx (1+\zeta) dW_{\theta}(t)$ for $\theta \approx 0$, and $ dU(t) \approx (1-\zeta) dW_{\theta}(t)$ for $\theta \approx \pi$, where $dW_{\theta}$ is a real-valued Wiener increment.

\begin{figure}[tp]
	\centering	\includegraphics[width=\columnwidth]{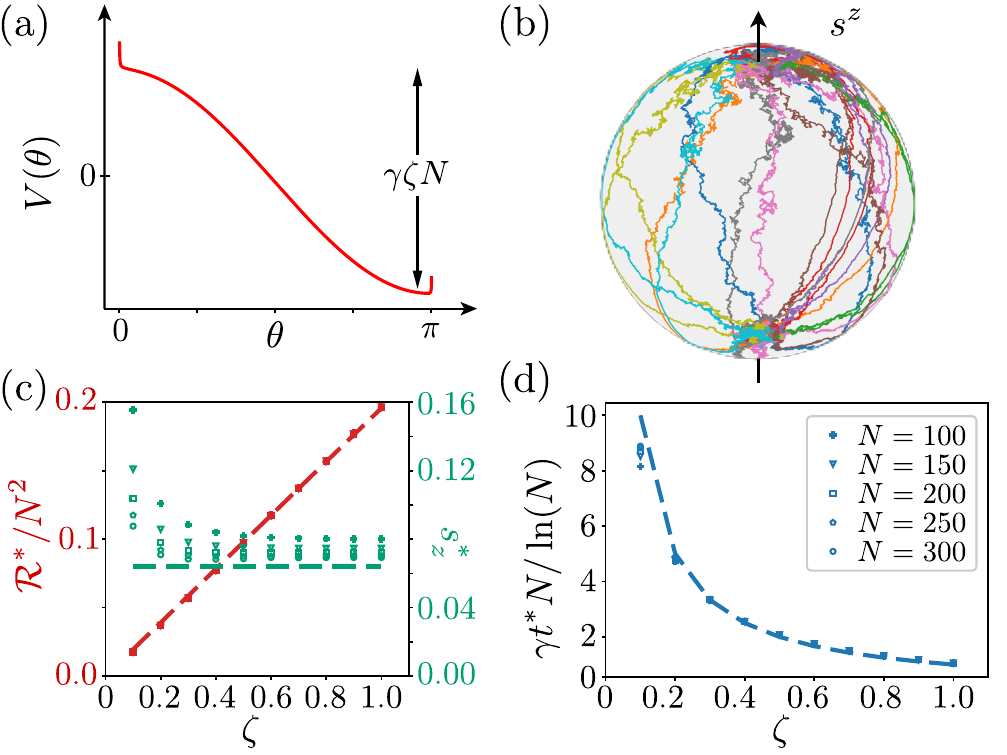}
	\caption{(a) Sketch of the effective potential $V(\theta)$ for describing time evolution of $\theta$, and (b) illustration of example trajectories for $\theta(t)$ and $\phi(t)$ on a Bloch sphere. For both plots, $N=100$ and $\zeta = 0.5$ have been used. The plots in (c) and (d) show the dependence of the maximal rate $\mathcal{R}^{*}/N^2$, the peak-magnetization, $s_{z}^{*}$, and the time of the peak, $\gamma t^{*} N/\ln(N)$, as a function of $\zeta$. The results indicated by symbols have been obtained from exact simulations of Eq.~\eqref{SSR_ME} for increasing $N$ using {\tt QuTiP} \cite{Qutip2012}. The dashed lines represent the theoretical predictions in Eq.~\eqref{MFRstar0} and Eq.~\eqref{MFtstar0}. The numerical data converges to our theoretical predictions as $\zeta N$ increases.}
	\label{Fig_Classical}
\end{figure}

From Eq.~\eqref{ThetaEOMmain} and the characteristic shape of $V(\theta)$ depicted in Fig.\,\ref{Fig_Classical}(a), we see that the superradiant decay occurs in three stages. For very small $\theta$, the initial dynamics is described by a Bessel process, $d\theta \approx  \frac{\gamma}{4} (\zeta+1)^2 \frac{1}{\theta} dt + \sqrt{\frac{\gamma}{2}} dW_{\theta}$,
which, starting from $\theta=0$, produces the distribution
\begin{equation}\label{Theta0initial}
    p(\theta,\Delta t) = \frac{2\theta}{\gamma (1+\zeta)^2 \Delta t} \exp{\left[-\frac{\theta^2}{\gamma (1+\zeta)^2 \Delta t}\right]}
\end{equation}
after a short time $\Delta t$. After a time of about $ \Delta t \approx 2/(\pi \gamma \zeta N)$, the collective term  $\sim N$ in Eq.~\eqref{Potential} starts to dominate and the system undergoes an approximately deterministic evolution with 
\begin{equation}\label{ThetaEOMSR}
    d\theta \approx \frac{\gamma}{2} (N-1) \zeta \sin\theta dt.
\end{equation}
It is important to note that this collective dynamics is independent of $\phi$ in spite of the squeezing, and thus leads to the same accelerated decay in all directions. This is in contrast to the single-particle case, where the correlations in the reservoir due to squeezing result in highly asymmetric decay rates~\cite{GardinerPRL1986}. 

In the last stage, the evolution of $\theta$ is stopped at a sharp wall at $\theta=\pi$. The diffusion terms become relevant again and give rise to an average value of $\langle\theta\rangle\neq \pi$ as $t\rightarrow \infty$, i.e., a nonvanishing residual excitation of the TLSs for $\zeta<1$. In the limit of large $N$, all three stages of this evolution can be evaluated analytically and we present the results in the End Matter. This allows us to make accurate predictions for the scaling of the superradiant peaks. Specifically, we obtain the maximal rate 
\begin{equation}\label{MFRstar0}
    \mathcal{R}^{*} \approx 0.195707 \times  \zeta  \gamma N^2,
\end{equation}
and the emission time 
\begin{equation}\label{MFtstar0}
    \gamma t^{*} = \frac{1}{\zeta}   \frac{\ln N}{N}  + O(1/N),
\end{equation}
and we also show that the average magnetization at the emission peak, $s_{z}^{*} = 2\braket{S_{z}}/N \approx 0.064$, is independent of the squeezing strength. For $\zeta =1$, we recover the known scaling relations for conventional Dicke superradiance \cite{DegiorgioOC1971,DegiorgioPRA1971,HaakePRA1972,GrossPhysRep1982,MalzPRA2022}. To verify the validity of these predictions together with the underlying assumption of purely classical trajectories, we compare them in Fig.~\ref{Fig_Classical}(c) and (d) with results obtained from exact simulations of Eq.~\eqref{SSR_ME}. We see that the numerical results converge to our analytic predictions as the parameter $\zeta N$ increases.

\emph{Super- and subradiance.}---In a second example, going beyond the restrictive case of a permutation-invariant ensemble, we study an array of TLSs located at positions $z_j$ along a 1D waveguide. This scenario can be modeled by the master equation (see, for example,~\cite{DzsotjanPRB2010,DzsotjanPRB2011,ZhengPRL2013,LalumierePRA2013})
\begin{equation}\label{SSR_ME_waveguide}
\frac{d}{dt} \rho  = -i[ H,\rho]+  \mathcal{D}[J_{\text{R}}] \rho +  \mathcal{D}[J_{\text{L}}] \rho ,     
\end{equation}
where $J_{\text{R/L}} = \sqrt{ \frac{\gamma}{2} } \sum_{j} e^{\mp i k_{0} z_{j}} \sigma_{j}^{-}$ are two jump operators for emission into right- and left-propagating modes with resonant wavevector $k_0$.  In Eq.~\eqref{SSR_ME_waveguide} the Hamiltonian $H = \frac{\gamma}{2} \sum_{j,l\neq j} \sin(k_{0} |z_{j} - z_{l}|) \sigma_{j}^{+} \sigma_{l}^{-}$ accounts for coherent photon-mediated interactions. Note that in the absence of any symmetries, exact simulations of Eq.~\eqref{SSR_ME_waveguide} are essentially impossible for large $N$.

\begin{figure}[tp]
	\centering	\includegraphics[width=\columnwidth]{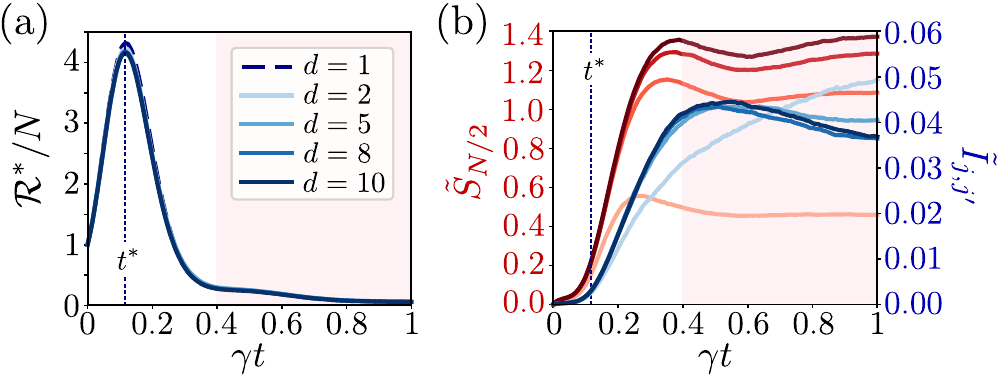}
	\caption{Results from a numerical QSD simulation of the decay of an ensemble of $N=50$ TLSs with spacing $k_0z_j =0.2\pi j$. The plots show (a) the time evolution of $\mathcal{R}/N$ and (b) the time evolution of average entanglement $\tilde{S}_{N/2}$ and the average mutual information $\tilde{I}_{j,j^{\prime}}$ with $j=N/2$ and $j^{\prime}=N/2+1$. Data for other pairs are shown in End Matter. The vertical solid line marks the time of the superradiant peak, and the pink shaded region indicates the regime, where the decay dynamics is dominated by subradiant states. The parameters for all simulations are  $N_{\text{tr}} =10^{3}$ and $\delta t = 2\times 10^{-4}/\gamma$.}
	\label{FigWQED}
\end{figure}

Generalizing the QSD approach to this setting~\cite{SupMat}, we simulate the the collective decay of $N=50$ TLSs into the waveguide, shown in Fig.~\ref{FigWQED}. For these simulations we consider an equally spaced array with position $k_0z_j =0.2\pi j$ that are incommensurate with the wavelength of the emitted light. Note that in this case, there are many collective subradiant states with strongly suppressed emission rates (see, for example,~\cite{AsenjoGarciaPRA2017,AsenjoGarciaPRX2017,AlbrechtNJP2019,HolzingerPRL2022,FasserOQ24,TiranovScience2023,SheremetRMP2023}), which can be populated during the decay process and thus determine the slow decay of the ensemble at long times. Indeed, from our simulations we see that after the superradiant peak, $\mathcal{R}$ has a long tail, which is a clear signature of subradiance. This becomes more evident in the average entanglement $\tilde{S}_{N/2}$, which is small around the superradiant peak but goes to a finite value in the subradiant regime, consistent with the appearance of entangled subradiant states. However, again, we see that for 2-local observables such as $\mathcal{R}$, product states ($d=1$) approximate both the super- and subradiant regime well. For non-local quantities such as $\tilde{S}_{N/2}$, a bond dimension $d\approx 10$ is required to reach  convergence. Like in squeezed superradiance, the average mutual information $\tilde{I}_{j,j^{\prime}}$ stays very small ($\lesssim  0.05$) even in the subradiant regime. Therefore, the structure in Eq.~\eqref{unravelrho12} is still valid in our unraveling, which makes the semiclassical theory ($d=1$) accurate for computing local observables.

\emph{Conclusion and outlook.}---In summary, we have shown that with a QSD unraveling, local observables in both super- and subradiant decay processes can be accurately computed using a mean-field approximation for each trajectory, even if a considerable amount of entanglement is generated in this process. The key is that the average mutual information $\tilde{I}$ between any pair of TLSs vanishes in the large-$N$ limit. Based on these insights, we provide a semiclassical theory for squeezed superradiance, which is exact for large $N$ and provides intuitive understanding. For more general settings, this provides a deeper theoretical justification for the use of semiclassical methods such as the discrete truncated Wigner approximation~\cite{HUberPRA2022,MinkSciPostPhys2023} or cumulant expansions~\cite{RubiesBigordaPRR2023} for simulating superradiance. Note, however, that by using an MPS representation, our QSD approach can systematically extend and benchmark such approximation schemes by increasing the bond dimension. 

More generally, entanglement can be a measure of computational complexity for simulating quantum systems on a classical computer. Since unraveling open system dynamics can drastically change the amount of observed entanglement~\cite{SkinnerPRX2019,LiPRB2019,CaoArxiv2019,VasseurPRB2019,VovkPRL2022,VovkPRA2024}), 
finding a suitable unraveling can sometimes greatly reduce computational complexity. The success of our numerical approach based on quantum state diffusion and matrix-product state suggests that it may also find applications in the simulation of other open quantum many-body systems.

\emph{Note added.}---During the completion of our work, two related articles show the absence of entanglement in conventional Dicke superradiance, i.e., in the permutation-invariant setting and without squeezing~\cite{BasslerArxiv2025,Rosario2025Arxiv}. Ref.~\cite{BasslerArxiv2025} presents a mathematical proof that there is no entanglement in conventional Dicke superradiance and Ref.~\cite{Rosario2025Arxiv} introduces a numerical unraveling scheme that is different from ours.

\emph{Acknowledgment.}---We thank Alexander Poddubny and Darrick Chang for stimulating discussions.
The numerical computation using MPS was implemented using {\tt ITensor.jl}~\cite{FishmanSciPostPhysCode2022}. We acknowledge support by the Deutsche Forschungsgemeinschaft (DFG, German Research Foundation)–522216022 and by the ERC project `ConsQuanDyn' (No.~851161). D.~M.\ acknowledges support
from the Novo Nordisk Foundation under grant numbers NNF22OC0071934 and NNF20OC0059939. This research is part of the Munich Quantum Valley, which is supported by the Bavarian state government with funds from the Hightech Agenda Bayern Plus.

\emph{Data availability.}---Numerical data for the figures are available at \cite{Zhang_data_2025}.

\bibliography{ContinuousMonitoringSuperradiance,BibFootnotes}

\clearpage

\newpage
\section{End Matter}

\begin{figure}\label{Fig_MoreDataPlot}
	\center
\includegraphics[width=\columnwidth]{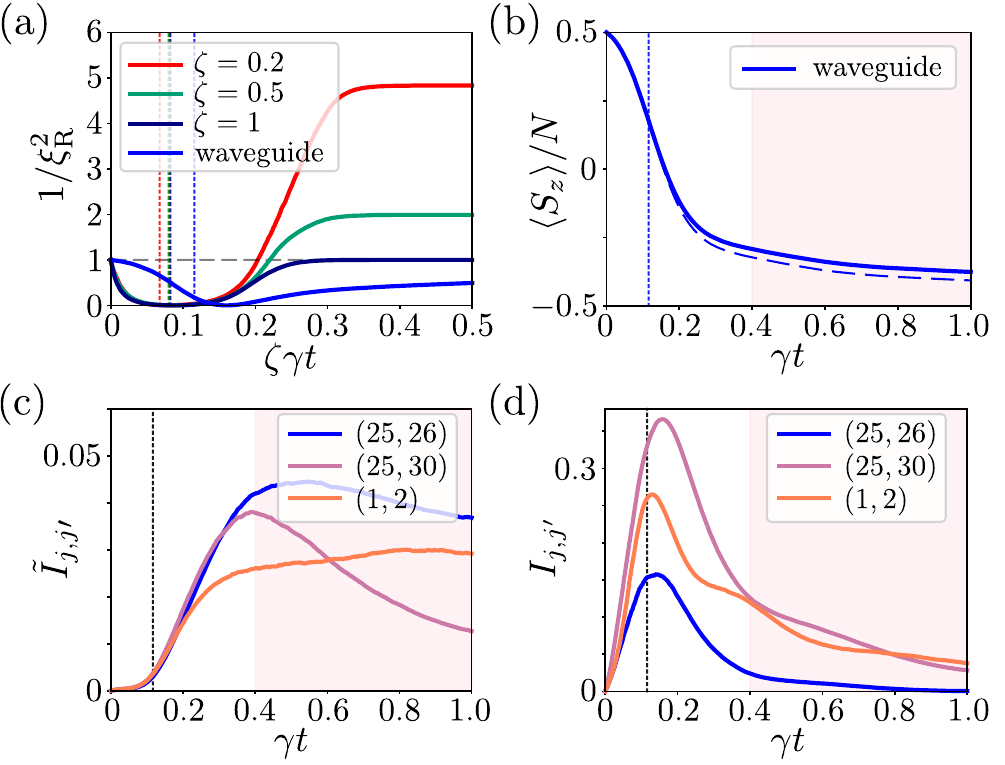}
	\caption{Additional data. (a) Time evolution of the inverse squeezing parameter, $1/\xi_{\text{R}}^{2}$, for squeezed Dicke superradiance and  for TLSs coupled to a 1D waveguide. (b) Plot of the total magnetization $\braket{S_{z}}/N$ for the waveguide setup. The dashed line represents the results obtained from MF trajectories. Time evolution of (c) $\tilde{I}_{j,j^{\prime}}$ and (d) $I_{j,j^{\prime}}$ for different pairs of TLSs for the waveguide setup. In all plots, $N=50$ TLSs and the vertical dotted line marks the position of the maximal decay rate. Other parameters are as in Fig.~\ref{Fig_PopRate} and Fig.~\ref{FigWQED}.}
\end{figure}

\emph{Derivation of the analytical theory for squeezed superradiance.}---Since our numerical results suggest that a mean-field (MF) theory is sufficient to describe local observables in this unraveling, here we analyze Eq.~\eqref{QSDgeneral} with the ansatz that $\ket{\psi}_{s}$ are product states. We can then derive MF equations of motion (EOM) for the spin variables. For an operator $O$, and when generalized to multiple decay channels with jump operators $J_k$, Eq.~\eqref{QSDgeneral} gives  
\begin{align}\label{QSDHeisenberg}
    d \braket{O} = &i \braket{ [H, O]} dt + \sum_{k} \braket{ J_{k}^{\dagger} O J_{k} - \frac{1}{2} \{O, J_{k}^{\dagger}J_{k} \} } dt \nonumber \\
    & + dW_{k} ( \braket{O J_{k}} - \braket{O} \braket{J_{k}} ) +  {\text{c.c.}}
\end{align}
In the permutation-invariant scenario with a single jump operator, $\braket{\sigma_{j}^{\alpha}}$ is independent of $j$ and we denote it as $s^{\alpha}$. Then, using the MF approximation 
$\braket{\sigma_{j}^{\alpha}\sigma_{j^{\prime}}^{\beta}} \approx \braket{\sigma_{j}^{\alpha}} \braket{\sigma_{j^{\prime}}^{\beta}}=s^\alpha s^\beta$ for $ j\neq j^{\prime}$, the EOM can be written as
\begin{align}\label{SqueezedSEOM}
    ds^{x} = &-\frac{\gamma}{2} \zeta^2 s^{x} dt + \frac{\gamma}{2} \zeta (N-1) s^{z}s^{x} dt \nonumber\\
    &+ \sqrt{\frac{\gamma}{2}} (\zeta s^{z}+1-s^{x}s^{x}) dW_{a} + \sqrt{\frac{\gamma}{2}} \zeta s^{x}s^{y} dW_{b}, \nonumber\\
    ds^{y} = &-\frac{\gamma}{2} s^{y} dt + \frac{\gamma}{2} \zeta (N-1) s^{z}s^{y} dt \\
    &- \sqrt{\frac{\gamma}{2}} (s^{z}+
    \zeta - \zeta s^{y}s^{y}) dW_{b} - \sqrt{\frac{\gamma}{2}} s^{x}s^{y} dW_{a}, \nonumber\\
    ds^{z} = &-\frac{\gamma}{2} [(\zeta^2+1)s^{z}+2\zeta] dt - \frac{\gamma}{2} \zeta (N-1) (s^{x} s^{x} + s^{y}s^{y}) \nonumber\\
    &- \sqrt{\frac{\gamma}{2}} (\zeta+s^{z}) s^{x} dW_{a} + \sqrt{\frac{\gamma}{2}} (1+\zeta s^{z}) s^{y} dW_{b}, \nonumber
\end{align}
where $dW_{a}$ and $dW_{b}$ are two real-valued Wiener increments. We can see that there are linear terms that describe the decay of single spins, and that there are nonlinear terms that describe spin-spin interactions, which are proportional to $\zeta(N-1)$. A value of $\zeta < 1$ affects the linear terms for $s^{x}$ but not for $s^{y}$, such that $s^{x}$ decays slower. However, the nonlinear term $\propto \zeta (N-1)$, which describes the collective decay, is the same for $s^{x}$ and $s^{y}$. For the waveguide setting with arbitrary positions, the corresponding mean-field equations are summarized in~\cite{SupMat}.

Since the trajectories are pure, the spin length is conserved and thus the spin variables live on the Bloch sphere [see Fig.\,\ref{Fig_Classical}(b)]. Then, by parameterizing $s^{x} = \sin \theta \sin \phi$, $s^{y} = \sin \theta \cos \phi$, and $s^{z} = \cos \theta $, we obtain the EOM for $\theta$ and $\phi$,
\begin{widetext}
\begin{equation}\label{ThetaPhiEOM}
\begin{split}
    d\theta ={}& \frac{\gamma}{2} \frac{1}{\sin\theta} \big[ (\zeta^2+1)\cos\theta + 2\zeta \big]dt - \frac{\gamma}{4} \frac{\cos\theta}{\sin\theta} \big[ \cos^2\phi (\zeta +\cos\theta)^2 + \sin^2\phi (1+\zeta\cos\theta)^2 \big]dt\\
    &+ \frac{\gamma}{2} (N-1) \zeta \sin\theta dt +\sqrt{\frac{\gamma}{2}} \cos\phi (\zeta + \cos\theta) dW_a - \sqrt{\frac{\gamma}{2}} \sin\phi (1+\zeta\cos\theta) dW_b,\\
     d\phi ={}& -\frac{\gamma}{2} \frac{1}{\sin^2\theta} \cos\phi \sin\phi \big[ (1+\zeta\cos\theta)^2-(\zeta + \cos\theta)^2 \big]dt \\
    &- \sqrt{\frac{\gamma}{2}} \frac{1}{\sin\theta} \sin\phi (1+\zeta\cos\theta) dW_a - \sqrt{\frac{\gamma}{2}} \frac{1}{\sin\theta} \cos\phi (\zeta + \cos\theta) dW_b.
\end{split}
\end{equation}
\end{widetext}
Numerical simulations of these EOM are equivalent to our MPS simulations with $d=1$. However, in this parametrization, we can get an analytical understanding in the limit $\zeta N \gg 1$, which is always satisfied for large enough $N$. 

For our fully inverted initial state we have $\theta(0)=0$, and $\theta(t)$ remains small for short enough times. By expanding Eq.~\eqref{ThetaPhiEOM} for small $\theta$, we obtain
\begin{equation}\label{ThetaEOMinital}
    d\theta \approx \frac{\gamma}{2} (N-1)\zeta \theta dt + \frac{\gamma}{4} (\zeta+1)^2 \frac{1}{\theta} dt + \sqrt{\frac{\gamma}{2}} (1+\zeta) dW_{\theta},
\end{equation}
where $dW_{\theta} = \cos\phi\, dW_{a} - \sin\phi\, dW_b $. Note that after this expansion, dependence on $\phi$ is completely absorbed into the noise process. During the initial stage, where $\theta \ll (\zeta +1)/\sqrt{2N\zeta}$, we can further ignore the first term and approximate
\begin{equation}
d\theta \approx  \frac{\gamma}{4} (\zeta+1)^2 \frac{1}{\theta} dt + \sqrt{\frac{\gamma}{2}} dW_{\theta}.
\end{equation}
In terms of the rescaled variable $\theta/[\sqrt{\gamma/2} (\zeta +1)]$, this is a Bessel process \cite{KarlinTaly1981} of dimension 2 starting from 0,  i.e. $\text{BES}_{0}^{2}$, whose solution is known. The distribution of $\theta$ after a evolution time $\Delta t$ is given by Eq.~\eqref{Theta0initial} in the main text. Since for this distribution the expectation value of $\theta$ is $E[\theta]  =  \frac{1}{2} (\zeta+1) \sqrt{\pi \gamma  \Delta t}$, we require the condition $\Delta t \ll 2/(\pi \gamma \zeta N)$ to keep the first term of \eqref{ThetaEOMinital} negligible.

After this initial stochastic evolution, we reach a value $\theta_{\rm 0}$ that makes the collective term dominate, such that
\begin{equation}\label{ThetaEOMSR}
    d\theta \approx \frac{\gamma}{2} (N-1) \zeta \sin\theta dt
\end{equation}
during the successive evolution. This evolution is deterministic and given by simple MF theory since fluctuations are negligible. With initial condition $\theta_{\text{0}}$, its solution is given by ${\sin\theta}/{(1+\cos\theta)} =\exp{[(N-1)\gamma \zeta t/2]}  {\sin\theta_{\text{0}}}/{(1+\cos\theta_\text{0})}$.
By assuming that $\theta_{\text{0}} \ll 1$, we obtain
\begin{equation}
    \cos\theta(t)  \simeq \frac{1 - e^{(N-1)\zeta \gamma t} \theta_{\text{0}}^2/4}{1 + e^{(N-1)\zeta \gamma t}\theta^2_{\text{0}}/4}.
\end{equation}
Here, we further assume that the continuous transition from the initial Bessel process in Eq.~\eqref{ThetaEOMinital} to the deterministic dynamics in Eq.~\eqref{ThetaEOMSR} can be approximated by a two-stage process. This means that we average the initial value of $\theta_{0}$ using the initial distribution in Eq.~\eqref{Theta0initial}.

In polar coordinates, the decay rate
\begin{equation}
\begin{split}
\mathcal{R} = {}&\frac{\gamma}{4} N \left[  (\zeta^2+1) E[\cos\theta] + 2\zeta   \right] \\& + \frac{\gamma}{4} N(N-1) \zeta E[\sin^2\theta],
\end{split}
\end{equation}
is dominated by the second term for large $N$. Then, by computing $E[\sin^2\theta]$ and finding the time when it reaches a maximum, we obtain the value for the superradiant time $t^{*}$,
\begin{equation} \label{Exptstar}
     \frac{1}{4} \gamma (\zeta+1)^2 \Delta t e^{(N-1)\zeta\gamma t^{*}} \approx 1.390537. 
\end{equation}
With this result, we find
\begin{equation}\label{MFRstar}
    \mathcal{R}^{*} \approx \frac{\gamma}{4} N^2 \zeta E[\sin^2\theta] \approx 0.195707 \times  \zeta  \gamma N^2,
\end{equation}
which is independent of the precise value of $\Delta t$ and recovers the scaling in Dicke superradiance with $\zeta =1$ \cite{DegiorgioOC1971,DegiorgioPRA1971,HaakePRA1972,GrossPhysRep1982,MalzPRA2022}. We see that the decay rate is simply suppressed with the squeezing parameter $\zeta$ in squeezed superradiance. Similarly the average population at $t^{*}$ is given by
\begin{equation}\label{MFszstar}
s_{z}^{*} = \braket{\sigma^{z}} = E[\cos\theta] \approx 0.064,
\end{equation}
which, interestingly, does not depend on squeezing. For the time $t^{*}$, Eq.~\eqref{Exptstar} gives
\begin{equation}\label{MFtstar}
    \gamma t^{*} = \frac{1}{\zeta}   \frac{\ln N}{N}  + O(1/N),
\end{equation}
which also recovers the $\frac{1}{N} \ln (N)$ scaling in Dicke superradiance with $\zeta =1$. Note that here we cannot obtain an expression for the $O(1/N)$ term, because it depends on $\Delta t\sim 1/N$, which is not precisely defined in our two-stage approximation.

For dynamics after the superradiant burst, we can expand the EOM around $\theta =\pi$. With $\Theta = \pi -\theta$ we obtain a radial Ornstein–Uhlenbeck process \cite{GardinerZoller2004},
\begin{equation}\label{ThetaEOMfinal}
    d\Theta = -\frac{\gamma}{2} (N-1) \zeta \Theta dt + \frac{\gamma}{4} (1-\zeta)^2 \frac{dt}{\Theta} + (1-\zeta) \sqrt{\frac{\gamma}{2}} dW_{\theta}^{\prime},
\end{equation}
where $dW_{\theta}^{\prime} = \cos\phi dW_a + \sin\phi dW_b$.   Note that for $\zeta < 1$ the noise term does not vanish anymore, which leads to a finite steady state population $(2\braket{S_{z}}+N) \approx (1-\zeta)^2/(2\zeta)$.

In summary, from the three stages of the dynamics given by Eq.\,\eqref{ThetaEOMinital}, \eqref{ThetaEOMSR}, and \eqref{ThetaEOMfinal}, we can describe the deterministic evolution of $\theta$ with an effective potential $V(\theta)$ as shown in Eq.\,\eqref{Potential}.
This leads to a very intuitive understanding of superradiance---$\theta$ flows from $0$ to $\pi$ with a maximum speed proportional to $\gamma \zeta N$ [see Fig.\,\ref{Fig_Classical}(a)], which translates into the $N^2$ scaling of the total decay rate \cite{SupMat}. By showing that the MF theory becomes exact for large $N$, we further validate the conclusion that the correlations between a pair of spins are fully unraveled with QSD as shown in Eq.~\eqref{unravelrho12}.

\emph{Spin squeezing parameter.}---The Wineland spin squeezing parameter $\xi_{\text{R}}^{2} $ is defined as~\cite{WinelandPRA1992}
\begin{equation}
    \xi_{\text{R}}^{2} = N \frac{\min (\Delta S_{\perp})^2}{\braket{S_{\vec{s}}}^2},
\end{equation}
where $\vec{s}$ is the direction of average angular momentum, and $\min (\Delta S_{\perp})^2$ is the minimum of angular momentum variance along directions perpendicular to $\vec{s}$. Notice that, due to the perforator $N$,  we cannot use the approximation in Eq.~\eqref{unravelrho12} to compute $\xi_{\text{R}}^{2}$. This is because, although corrections to the approximation in Eq.~\eqref{unravelrho12} vanish as $N\rightarrow\infty$, these corrections when multiplied by $N$ can converge to a nonzero value.

\emph{Additional data.}---In Fig.~\ref{Fig_MoreDataPlot}(a) we compare the squeezing parameter $\xi_{\text{R}}^{2}$ for squeezed superradiance with that of the waveguide setup. This parameter witnesses the steady-state entanglement for $\zeta <1$. However, we find that $\xi_{\text{R}}^{2}>1$ in the waveguide case, despite a large average entanglement that suggests that the subradiant states are entangled. Therefore, other entanglement witnesses must be employed to investigate this aspect further. In Fig.~\ref{Fig_MoreDataPlot}(b), we compare the decay of the magnetization $\braket{S_{z}}/N$ computed with MF theory with MPS simulations with $d=10$. This comparison again confirms that MF trajectories work well for local observables. Finally, in Fig.~\ref{Fig_MoreDataPlot}(c) and (d), we plot the average mutual information $\tilde{I}_{j,j^{\prime}}$ and the actual mutual information $I_{j,j^{\prime}}$ for different sets of qubit pairs. This plot demonstrates that also in the waveguide setup, the average mutual information is vanishingly small in this system across all pairs. Notice that, TLSs with phase difference $n \pi$ ($n\in\mathbb{Z}$) have larger correlations.


\newpage

\widetext

\clearpage

\setcounter{equation}{0}
\setcounter{figure}{0}
\setcounter{table}{0}
\setcounter{page}{1}
\makeatletter
\renewcommand{\theequation}{S\arabic{equation}}
\renewcommand{\thefigure}{S\arabic{figure}}
\renewcommand{\bibnumfmt}[1]{[S#1]}
\renewcommand{\citenumfont}[1]{S#1}

\begin{center}
	
	{\large\bf Supplemental Material for ``Unraveling superradiance: Entanglement and mutual information in collective decay''}
	
	\vspace{0.5cm}

	Xin  H.  H.  Zhang,$^{1,2,3}$ Daniel Malz$^{4}$ and Peter Rabl$^{1,2,3}$

    {\it 
$^{1}${Technical University of Munich, TUM School of Natural Sciences, Physics Department, 85748 Garching, Germany} \\
$^{2}${Walther-Mei{\ss}ner-Institut, Bayerische Akademie der Wissenschaften, 85748 Garching, Germany}\\
$^{3}${Munich Center for Quantum Science and Technology (MCQST), 80799 Munich, Germany} \\
$^{4}${Department of Mathematical Sciences, University of Copenhagen, Universitetsparken 5, 2100 Copenhagen, Denmark}
}
\end{center}

\subsection{(i) Other parameterization of squeezed superradiance}

The jump operators in squeezed superradiance [Eq.~(1) in the main text] can be expressed in a number of different ways. They are mathematically identical, but choosing one or the other parametrization may be more useful for a given experimental setup. We present here three additional equivalent ways of express the jump operator and the corresponding expressions for the maximal rate and the time of the superradiant peak:
\begin{itemize}
	\item[(i)] $J = \sqrt{\Gamma} (\cos\varphi S^{-} + \sin\varphi S^{+} ) $ with $S^{\pm} = \sum_{j} \sigma^{\pm}$, and $\varphi \in [0, \pi/4]$.\newline
	Then we can identify $\sqrt{\gamma} (1+\zeta)/2 =\sqrt{\Gamma} \cos\varphi $ and $\sqrt{\gamma} (1-\zeta)/2 = \sqrt{\Gamma} \sin\varphi$, we get $\zeta = (\cos\varphi - \sin\varphi)/(\cos\varphi + \sin\varphi)$ and $\gamma = (1 + 2 \cos\varphi \sin\varphi)\Gamma $. Therefore, 
	\begin{equation}
		\mathcal{R}^{*}  \approx 0.2 (\cos^2\varphi - \sin^2\varphi ) \Gamma N^2 ,\qquad
		t^{*}  =  \frac{1 }{\cos^2\varphi - \sin^2\varphi } \frac{1}{\Gamma}  \frac{\ln N}{N}. 
	\end{equation}
	
	\item[(ii)] $J = \sqrt{\Gamma} [ \cosh(r) S^{-} + \sinh(r) S^{+} ] $ with $r\in [0,\infty)$.\newline
	Setting $\zeta = e^{-2 r}$ and $\gamma = \Gamma e^{2 r}$ yields
	\begin{equation}
		\mathcal{R}^{*}  \approx 0.2  \Gamma N^2 ,\qquad
		t^{*}  =   \frac{1}{\Gamma}  \frac{\ln N}{N}. 
	\end{equation}
	Interestingly, these quantities do not depend on $r$. It is a convenient choice to keep the decay dynamics invariant while changing the degree of spin squeezing.
	
	\item[(iii)] $J = \sqrt{\Gamma} \big( S^{-} + \mathcal{E} S^{+} \big) $ with $ \mathcal{E} \in [0,1]$.\newline
	We get $\zeta = (1-\mathcal{E})/(1+\mathcal{E})$ and $\gamma = \Gamma (1+\mathcal{E})^2$, and thus
	\begin{equation}
		\mathcal{R}^{*}  \approx 0.2  \Gamma (1-\mathcal{E}^2) N^2 ,\qquad
		t^{*} =   \frac{1}{\Gamma (1-\mathcal{E}^2)}  \frac{\ln N}{N}. 
	\end{equation}
\end{itemize}


\subsection{(ii) QSD+MF for waveguide QED}
Here we derive the EOM of spin variables $s_{j}^{\alpha} = \braket{\sigma_{j}^{\alpha}}$ for an array of TLSs coupled with a 1D waveguide. Applying the MF approximation to Eq.~(11) in the main text, we obtain
\begin{equation}\label{spinEOMstep2}
	\begin{split}
		\frac{d}{dt} s_{j}^{x} =& - \frac{\gamma}{2}s_{j}^{x} + \frac{\gamma}{2} \sum_{l \neq j} \sin(k_{0} |z_{l} - z_{j}|) s_{l}^{y} s_{j}^{z}  +  \frac{\gamma}{2} \sum_{l\neq j} \cos(k_{0} (z_{j}-z_{l})) s_{j}^{z}  s_{l}^{x} 
		\\& + \sqrt{\frac{\gamma}{2}} \frac{1}{2} ( s_{j}^{z} +1 - s_{j}^{x} s_{j}^{x} +i s_{j}^{x} s_{j}^{y} ) ( e^{-i k_{0} z_{j}} dW_1 + e^{i k_{0} z_{j}} dW_2 ) + \text{c.c.} , \\ 
		\frac{d}{dt} s_{j}^{y} =&  - \frac{\gamma}{2}s_{j}^{y} -\frac{\gamma}{2} \sum_{l \neq j} \sin(k_{0} |z_{l} - z_{j}|) s_{l}^{x} s_{j}^{z}  +  \frac{\gamma}{2} \sum_{l\neq j} \cos(k_{0} (z_{j}-z_{l})) s_{j}^{z}  s_{l}^{y} 
		\\&+ \sqrt{\frac{\gamma}{2}} \frac{1}{2} ( -i  s_{j}^{z} -i - s_{j}^{y} s_{j}^{x} +i s_{j}^{y} s_{j}^{y} ) ( e^{-i k_{0} z_{j}} dW_1 + e^{i k_{0} z_{j}} dW_2 ) + \text{c.c.} , \\
		\frac{d}{dt} s_{j}^{z} =&  -\gamma (1+ s_{j}^{z}) + \frac{\gamma}{2} \sum_{l \neq j} \sin(k_{0} |z_{l} - z_{j}|) ( s_{l}^{x} s_{j}^{y} - s_{l}^{y} s_{j}^{x} )  - \frac{\gamma}{2} \sum_{l\neq j} \cos(k_{0} (z_{j}-z_{l})) ( s_{j}^{x}  s_{l}^{x} + s_{j}^{y}  s_{l}^{y})  \\&   - \sqrt{\frac{\gamma}{2}} \frac{1}{2} ( 1+ s_{j}^{z}) ( s_{j}^{x} - i s_{j}^{y}) ( e^{-i k_{0} z_{j}} dW_1 + e^{i k_{0} z_{j}} dW_2 ) + \text{c.c.} ,
	\end{split}
\end{equation}
where $dW_1$ and $dW_2$ two complex Wiener processes associated with the two decay channels. Numerical simulation of Eq.~\eqref{spinEOMstep2} corresponds to QSD+MPS simulation with $d=1$. The local reduced density matrices are constructed from the correlators $\braket{\sigma_{j}^{\alpha}\sigma_{j^{\prime}}^{\beta}} \approx E[ s_{j}^{\alpha}s_{j^{\prime}}^{\beta} ]$.

\end{document}